\documentstyle[aps,graphicx,colordvi]{revtex}

\begin{document}
\centerline{Date: \today\hfill{}}

\begin{center}
{\large Phenomenology of Radion in Randall-Sundrum Scenario}
\vskip 0.7cm
Kingman Cheung \\
{\it Department of Physics, University of California, Davis, CA 95616 USA} \\
{\it National Center for Theoretical Science, National Tsing Hua University,\\
Hsinchu, Taiwan R.O.C.}
\footnote{Present address.\\
  Email address: {\tt cheung@phys.cts.nthu.edu.tw}}
\\
\end{center}

\begin{abstract}
The success of the Randall-Sundrum scenario relies on  
stabilization of the modulus field or the radion, which is the scalar field
about the background geometry.
The stabilization mechanism proposed by Goldberger and Wise has 
the consequence that this radion is lighter than the graviton
Kaluza-Klein states so that the first particle to be discovered is the 
radion.  In this work, we study in details the decay, production, and
detection of the radion at hadronic, $e^+ e^-$, and $\gamma\gamma$ colliders.
\end{abstract}

\section{Introduction}

The standard model (SM) has been very successful in accounting for almost 
all experimental data.
Nevertheless, the SM can only be a low energy limit of
a more fundamental theory because it cannot explain
a number of theoretical issues, one of which is the gauge hierarchy problem 
between the only two known scales in particle physics -- the weak and Planck 
scales.  This longstanding problem has been the subject of attention for a 
long time, such as supersymmetric and technicolor theories.

Recent advances in string theories have revolutionized our perspectives
and understanding of the problems, namely, the Planck, grand unification, and
string scales can be brought down to a TeV range with the help of extra 
dimensions, compactified or not.   Arkani-Hamed {\it et al.} \cite{add}
proposed that using compactified dimensions of large size (as large as mm)
 can bring the
Planck scale down to TeV range.  In this model, the SM fields reside on a 
3-brane while only gravity is allowed in the extra dimensions.  The 
phenomenology is due to the Kaluza-Klein states of the graviton in the bulk. A 
summary of collider phenomenology can be found in Ref. \cite{me-pascos}.
Another interesting model was proposed by Dienes {\it et al.} \cite{keith}
that with gauge fields and/or fermion fields allowed in the extra dimensions,
the grand unification scale is brought down to TeV range.  

Randall and Sundrum (RS) \cite{RS} proposed a 5-dimensional space-time 
model with a 
nonfactorizable metric to solve the hierarchy problem.  The extra dimension is
a single $S^1/Z_2$ orbifold, in which two 3-branes of opposite tensions 
reside at the two fixed points (boundaries) of the orbifold and a 
cosmological constant in the bulk serves as the source
 for 5-dimensional gravity.
The resulting space-time metric is nonfactorizable and depends on the radius
$r_c$ of the extra dimension
\begin{displaymath}
ds^2 = e^{-2k r_c |\phi|} \; \eta_{\mu\nu} dx^\mu \, dx^\nu - r_c^2 d\phi^2\;,
\end{displaymath}
where $k$ is a parameter of order of $M$, the effective Planck scale of the
theory, and $\phi$ is the coordinate for the extra dimension.  The two 
3-branes reside at $\phi=0$ and $\phi=\pi$.  The 4-dimensional Planck mass
$M_{\rm Pl}$ is related to the scale $M$ by
\begin{displaymath}
M_{\rm Pl}^2  = \frac{M^3}{k}\, \left( 1-e^{-2k r_c \pi} \right ) \;,
\end{displaymath}
so that $M_{\rm Pl}$ is of order $M$.  
Thus, a field that is confined to the 3-brane at $\phi=\pi$ with mass
parameter $m_0$ will have a physical mass of $m_0 e^{-k r_c \pi}$.  For
$k r_c$ around  12, the weak scale can be dynamically generated from the
fundamental scale $M$.

The RS model has a four-dimensional massless scalar,
the modulus or radion, about the background geometry:
\begin{displaymath}
ds^2 = e^{-2 k |\phi| T(x)} g_{\mu\nu}(x) \, dx^\mu dx^\nu - T^2(x) d\phi^2 \;,
\end{displaymath}
where $g_{\mu\nu}(x)$ is the 4-dimensional graviton and $T(x)$ is the radion.
The most important ingredients of the above brane configuration are the 
required 
size of the modulus field such that it generates the desired weak scale 
from the scale $M$ and the stabilization of the modulus field at this value.
A stabilization mechanism was proposed by Goldberger and Wise \cite{GW1} 
that a bulk scalar field propagating in the background solution of the
metric can generate a potential that can stabilize the modulus field.  The
minimum of the potential can be arranged to give the desired value of 
$k r_c$ without fine tuning of parameters.
It has been shown \cite{GW2} that if a large value of $k r_c \sim 12$, needed
to solve the hierarchy problem, arises from a small bulk scalar mass then the
modulus potential near its minimum is nearly flat for values of modulus VEV
that solves the hierarchy problem.  As a consequence, besides getting a mass
the modulus field 
is likely to be lighter than any Kaluza-Klein modes of any bulk
field.  The mass of the radion is of order of $O(\rm TeV)$ and the strength
of coupling to the SM fields is of order of $O(1/{\rm TeV})$.
Therefore, the detection of this radion will be the first
signature of the RS model and the stabilization mechanism by Goldberger and
Wise.  Furthermore, the radion has to have mass in order for a brane world
(RS model here) to reproduce the ordinary 4-dimensional Einstein gravity
\cite{csaki}.

In the present paper, we perform a comprehensive study of the decay, 
production,
and detection and possible existing limits of the radion.  Similar works
have been considered in Refs. \cite{GRW,MD,BKLL}.  The improvement here
includes: (i) more production channels and cross sections are studied,
e.g., $p \bar p \to t\bar t \phi$ and $p\bar p \to W\phi,Z\phi$,
(ii) we also perform background studies on selected channels and their
feasibility of detection, and (iii) we check against the existing data on
the Higgs boson at the Tevation and at the LEP to see if the data can place 
some limits on the radion.
 
The organization of the paper is as follows.  In the next section, we summarize
the effective interactions of the radion. In Sec. III, we present the decay
of the radion.  In Sec. IV, we calculate the production cross sections for 
all relevant channels at $e^+ e^-$, hadronic, and photon colliders. 
Section V consists of background studies and detection feasibility of selected
channels.  We conclude in Sec. VI.

\section{Effective Interactions}

The interactions of the radion with the SM particles on the brane are 
model-independent and are governed by 4-dimensional general covariance, 
and thus given by this Lagrangian
\begin{equation}
\label{T}
{\cal L}_{\rm int} = \frac{\phi}{\Lambda_\phi} \; T^\mu_\mu ({\rm SM}) \;,
\end{equation}
where $\Lambda_\phi= \langle \phi \rangle$ is of order TeV, and $T_\mu^\mu$
is the trace of SM energy-momentum tensor, which is given by
\begin{equation}
T^\mu_\mu ({\rm SM}) = \sum_f m_f \bar f f - 2 m_W^2 W_\mu^+ W^{-\mu} 
-m_Z^2 Z_\mu Z^\mu + (2m_h^2 h^2 - \partial_\mu h \partial^\mu h  ) + ... \;,
\end{equation}
where $...$ denotes higher order terms.
The couplings of the radion with fermions and $W$, $Z$ and Higgs 
bosons are given in Eq. (\ref{T}).  

\begin{figure}[t]
\centering
\includegraphics[width=6.5in]{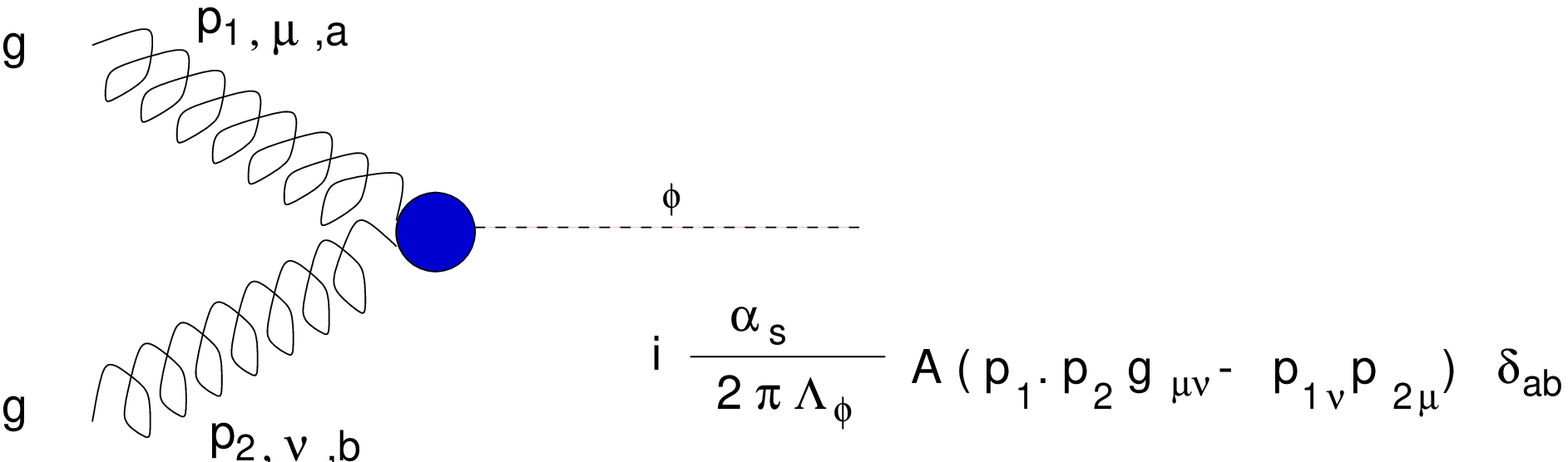}
\vskip1in
\includegraphics[width=6.5in]{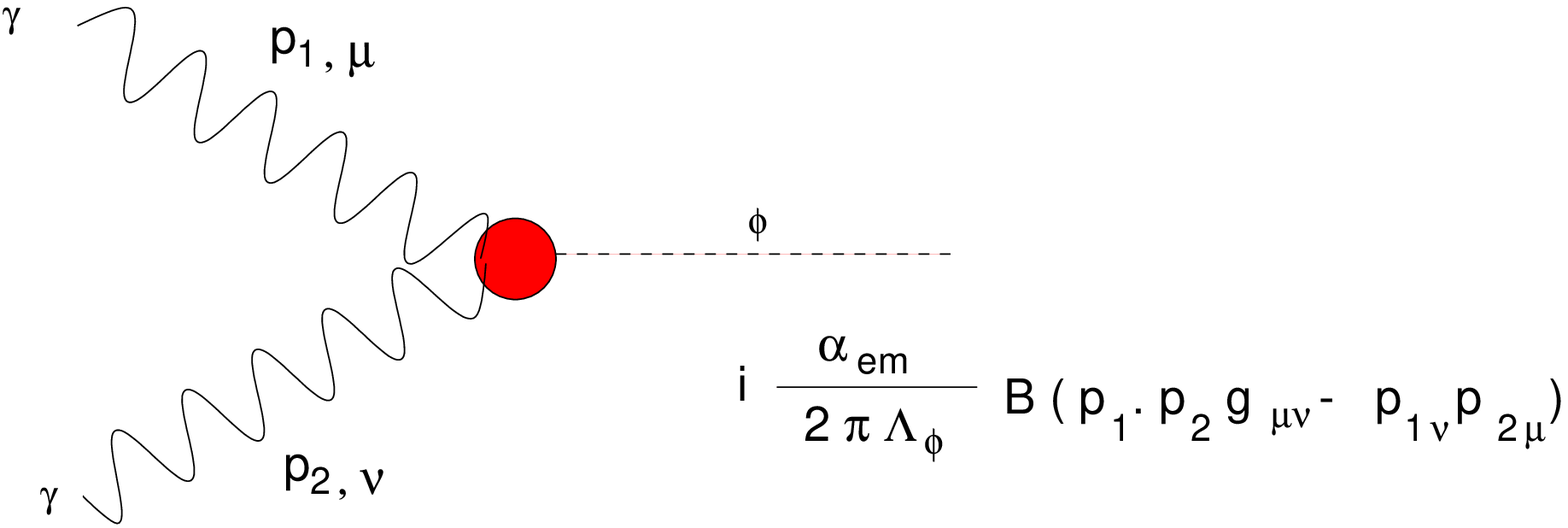}
\vskip0.1in
\caption{\small \label{fig1}
Feynman diagrams and rules for (a) $gg\phi$ vertex and (b) $\gamma\gamma\phi$
vertex.  The constant $A= b_{\rm QCD} + y_t(1+(1-y_t)f(y_t))$ and
$B=b_2+b_Y - (2+3y_W +3y_W(2-y_W)f(y_W) ) + \frac{8}{3}y_t (1+(1-y_t)f(y_t))
$, where $y_i = 4 m_i^2/2p_1 \cdot p_2$.
 $a$ and $b$ are color indices of the gluons.}
\end{figure}

For the coupling of the radion to a pair of gluons (photons), there are
contributions from 1-loop diagrams with the top-quark (top-quark 
and $W$) in the loop, and from the trace anomaly.
The contribution from the trace anomaly for gauge fields is given by
\begin{equation}
T^\mu_\mu({\rm SM})^{\rm anom} = \sum_a \frac{\beta_a (g_a)}{2g_a} 
F_{\mu\nu}^a F^{a \mu\nu} \;.
\end{equation}
For QCD $\beta_{\rm QCD}/2g_s = -(\alpha_s/8\pi) b_{\rm QCD}$, where
$b_{\rm QCD} = 11 - 2 n_f/3$ with $n_f=6$.  Thus, 
the effective coupling of $\phi g g$, including the 1-loop diagrams 
of top-quark 
and the trace anomaly contributions (Fig. \ref{fig1}(a)) is given by
\begin{equation}
\frac{i \delta_{ab}\alpha_s}{2\pi \Lambda_\phi}
\left[ b_{\rm QCD} + y_t ( 1+ (1-y_t)f(y_t) ) \right ]
\left( p_1 \cdot p_2 g_{\mu \nu} - p_{2_\mu} p_{1_\nu} \right ) \;,
\end{equation}
where $y_t= 4 m_t^2/2p_1 \cdot p_2$ and
the momentum and Lorentz index assignments are given in Fig. \ref{fig1}.

The effective coupling of $\phi \gamma\gamma$, including the 1-loop diagrams 
of the top-quark and $W$ boson, and the trace anomaly contributions 
(Fig. \ref{fig1}(b)) is given by
\begin{equation}
\frac{i \alpha_{\rm em}}{2\pi \Lambda_\phi}
\left[ b_2 + b_Y - (2+3y_W +3y_W (2-y_W)f(y_W) ) + \frac{8}{3} \,
y_t ( 1+ (1-y_t)f(y_t) ) \right ]
\left( p_1 \cdot p_2 g_{\mu \nu} - p_{2_\mu} p_{1_\nu} \right ) \;,
\end{equation}
where $y_i= 4 m_i^2/2p_1 \cdot p_2$, $b_2=19/6$ and $b_Y=-41/6$, and 
the momentum and Lorentz index assignments are given in Fig. \ref{fig1}.
In the above, the function $f(z)$ is given by
\begin{displaymath}
f(z) = \left \{ \begin{array}{cr}
\left[ \sin^{-1} \left(\frac{1}{\sqrt{z}} \right ) \right ]^2\;, & z \ge 1 \\
-\frac{1}{4} \left[ \log \frac{1+\sqrt{1-z}}{1-\sqrt{1-z}} - i \pi \right ]^2
\;, & z <1 
\end{array}
\right . \;.
\end{displaymath}
We now have all necessary couplings to perform calculations on decays and
production of the radion.

\section{Decays of Radion}

With the above interactions we can calculate the partial widths of the radion
into $gg,\gamma\gamma, f\bar f, WW, ZZ$, and $hh$. The partial widths are
given by
\begin{eqnarray}
\Gamma(\phi \to gg) &=& \frac{\alpha_s^2 m^3_\phi}{32 \pi^3 \Lambda_\phi^2}
 \left | b_{\rm QCD} + x_t( 1+ (1-x_t) f(x_t) ) \right |^2 \;, \\
\Gamma(\phi \to \gamma\gamma) &=& \frac{\alpha_{\rm em}^2 m^3_\phi}
{256 \pi^3 \Lambda_\phi^2}
\left | b_2+b_Y - (2+3 x_W+3x_W(2-x_W)f(x_W)) + \frac{8}{3} x_t( 1+ (1-x_t) 
f(x_t) ) \right |^2 \;, \\
\Gamma(\phi \to f\bar f) &=& \frac{N_c m_f^2 m_\phi}{8\pi \Lambda^2_\phi}
(1-x_f)^{3/2} \;, \\
\Gamma(\phi \to W^+ W^-) &=& \frac{m_\phi^3}{16 \pi \Lambda_\phi^2}\;
 \sqrt{1- x_W} \left( 1 - x_W +\frac{3}{4} x_W^2 \right ) \;, \\
\Gamma(\phi \to ZZ) &=& \frac{m_\phi^3}{32 \pi \Lambda_\phi^2}\;
 \sqrt{1- x_Z} \left( 1 - x_Z +\frac{3}{4} x_Z^2 \right ) \;, \\
\Gamma(\phi \to hh) &=& \frac{m_\phi^3}{32\pi \Lambda_\phi^2}\; 
  \sqrt{1-x_h} \left( 1+ \frac{x_h}{2} \right )^2 \;,
\end{eqnarray}
where $x_i = 4 m_i^2/m_\phi^2 (i=f,W,Z,h)$ and $N_c=3(1)$ for quarks (leptons).

\begin{figure}[t]
\centering
\includegraphics[width=6.5in]{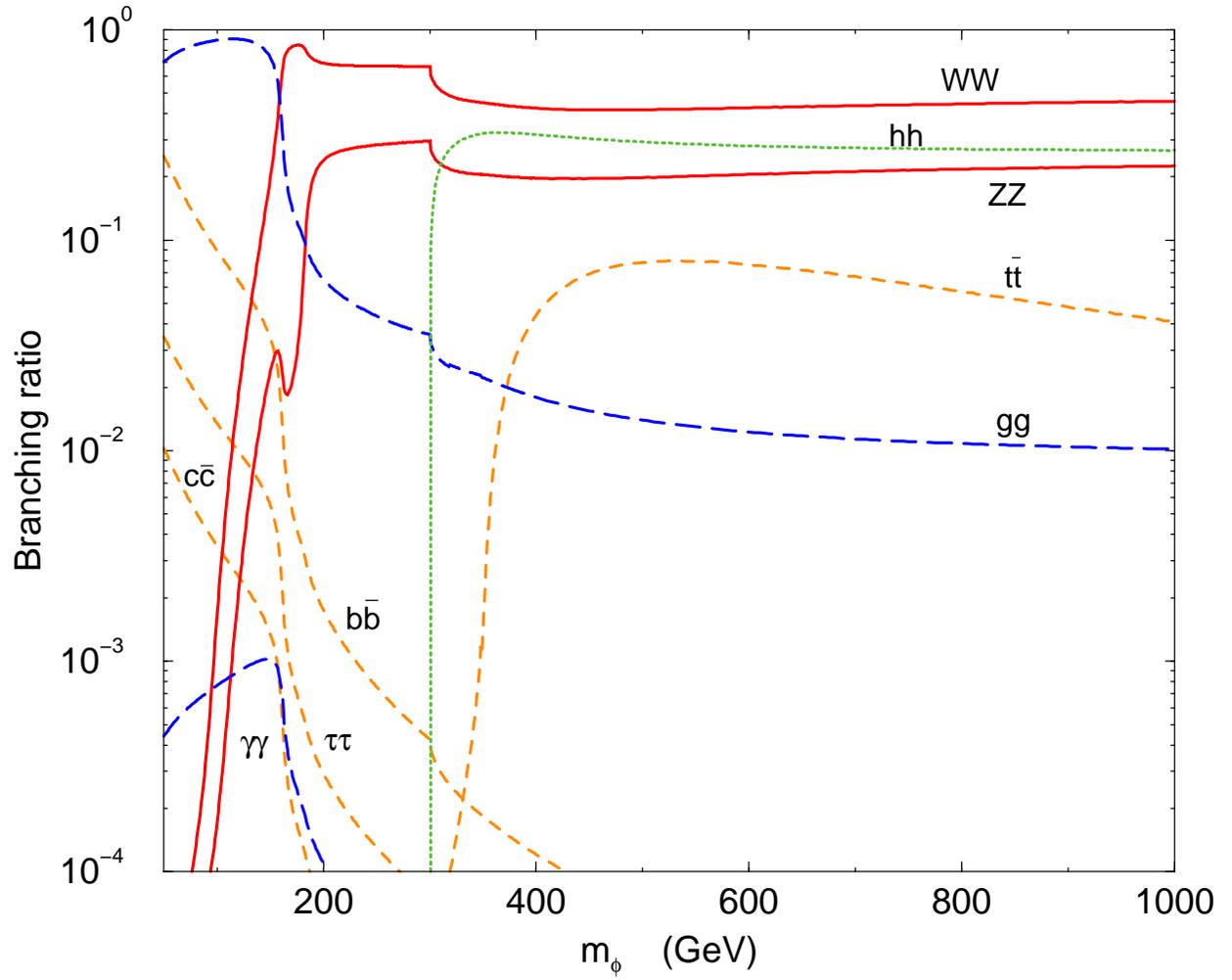}
\vskip0.2in
\caption{\label{decay} \small
Branching ratios of the radion versus $m_\phi$. Here we have used $m_h=150$ 
GeV.}
\end{figure}

In calculating the partial widths into fermions, we have used the three-loop
running masses with scale $Q^2 = m_\phi^2$.   
Figure \ref{decay} shows the branching ratios of the radion up to 1 TeV.
Note that the branching ratios are independent of $\Lambda_\phi$.
We have also allowed the off-shell decays of the $W$ and $Z$ bosons and 
that of the top quark, thus giving smooth onset of the $WW,ZZ$, and $t\bar t$
curves.  The features of decay branching ratios are
similar to the decay of the Higgs boson, except for as follows. 
At $m_\phi \alt 150$ GeV, the decay width is dominated by $\phi \to gg$ while 
the decay width of the SM Higgs boson is dominated by the 
$b\bar b$ mode.  At larger $m_\phi$, $\phi$ also decays into a pair of
Higgs bosons ($\phi \to hh$) if kinematically allowed while the SM
Higgs boson cannot.  Similar to the SM Higgs boson, 
as $m_\phi$ goes beyond the $WW$ and $ZZ$ 
thresholds, the $WW$ and $ZZ$ modes dominate with the $WW$ partial width 
about a factor of two of the $ZZ$ partial width.

\section{Production of Radion}

In the following, our discussions are divided according to
 different colliders, namely, hadron colliders (Tevatron and LHC), 
lepton colliders ($e^+ e^-, \mu^+ \mu^-$), and photon colliders.

\subsection{Hadronic colliders}

\begin{figure}[t]
\centering
\includegraphics[width=5.5in]{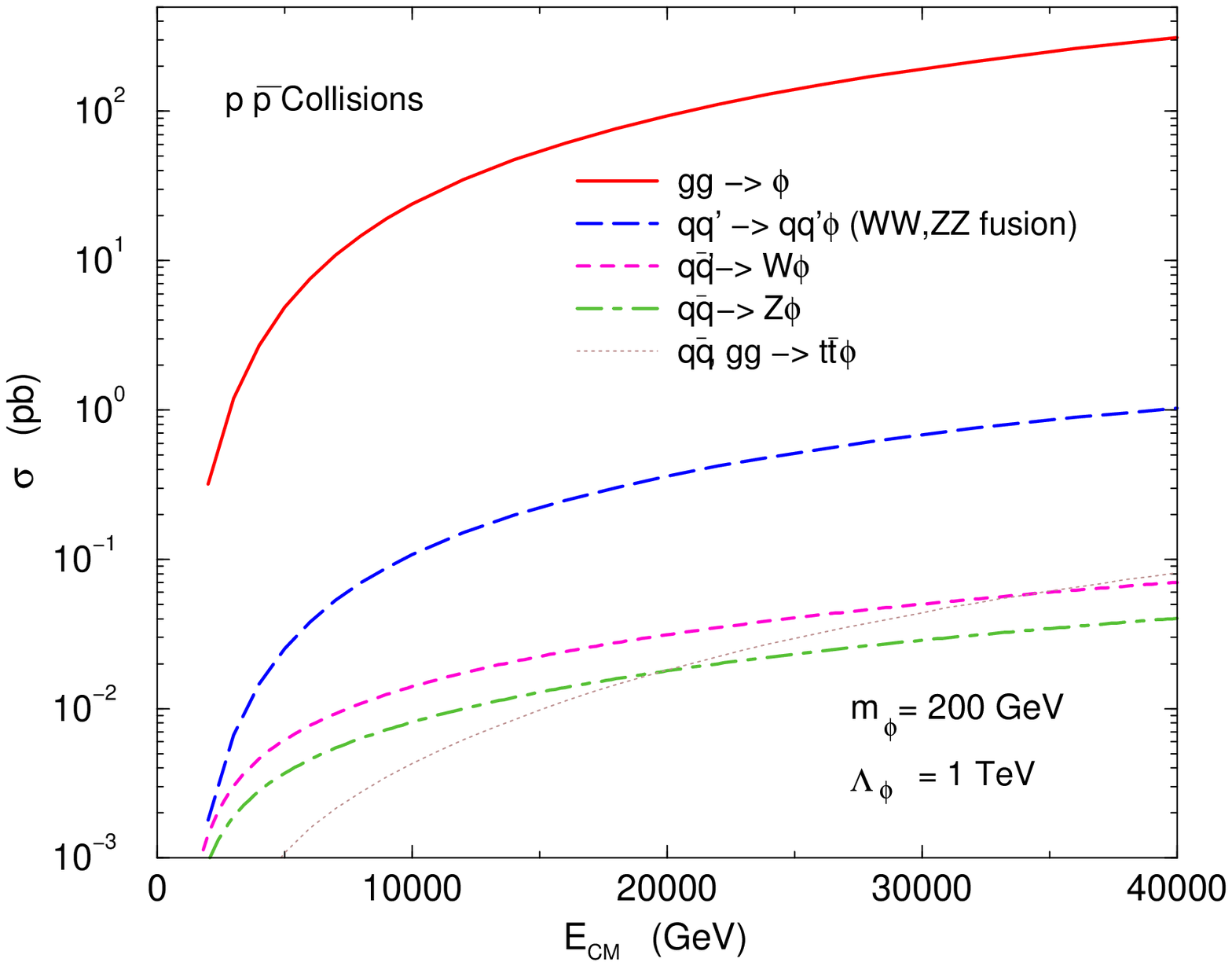}
\includegraphics[width=5.5in]{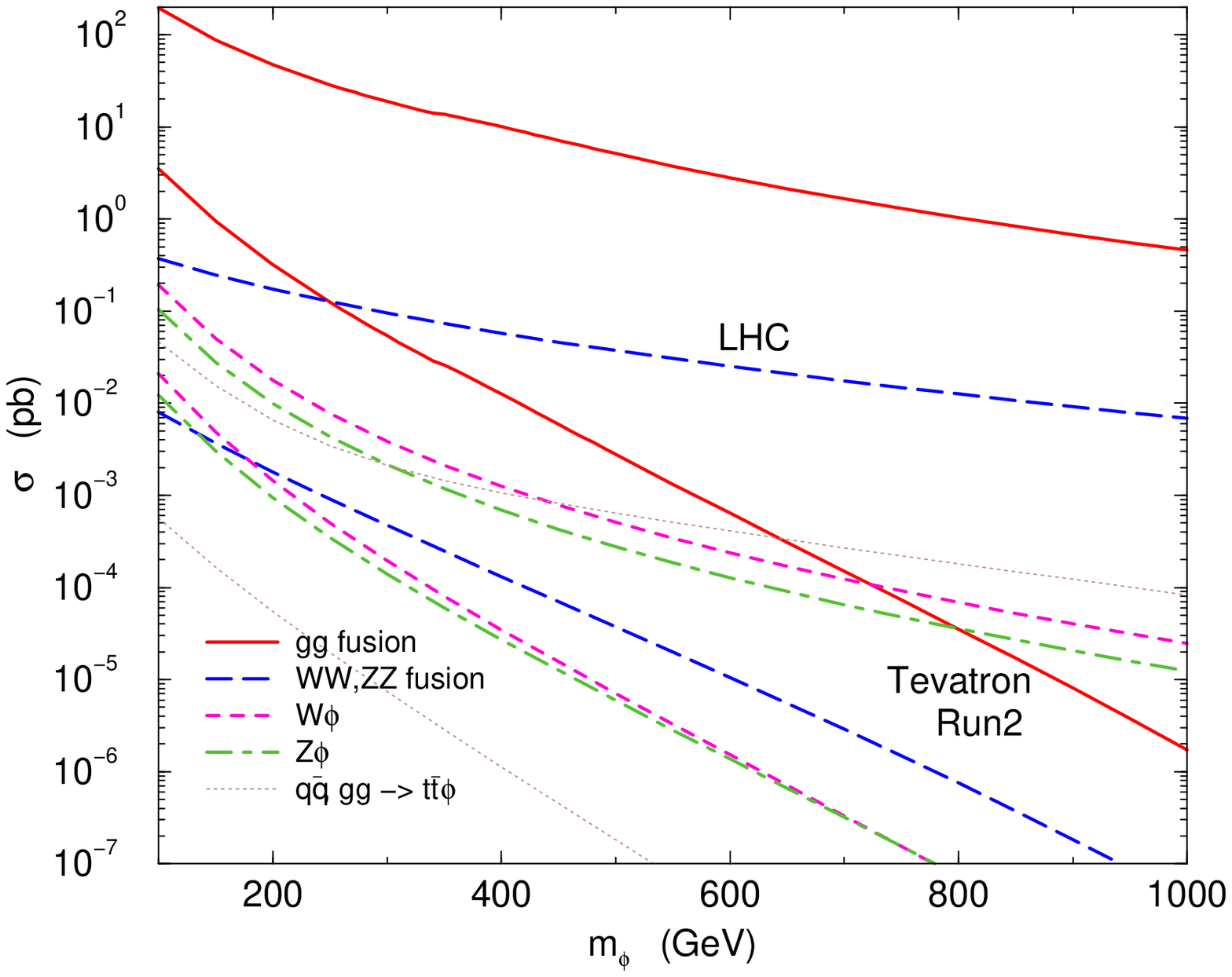}
\vskip0.1in
\caption{\label{pp-fig} \small
(a) Production cross sections versus the center-of-mass energy $E_{CM}$ for
$p\bar p \to \phi$ ($gg$ fusion), $p\bar p \to q q' \phi$ ($WW,ZZ$ fusion),
$p\bar p \to W\phi$, $p\bar p \to Z\phi$, and $p\bar p \to t \bar t \phi$.
(b) Production cross sections versus $m_\phi$.
}
\end{figure}

The production channels at hadronic colliders include
\begin{eqnarray}
gg &\to& \phi \nonumber\\
q \bar q' &\to& W \phi \nonumber \\
q \bar q &\to& Z \phi \nonumber\\
q q' &\to& q q' \phi \;\; \mbox{($WW,ZZ$ fusion)} \nonumber\\
q \bar q \;, \;\; gg &\to& t \bar t \phi \nonumber \;.
\end{eqnarray}
Similar to the SM Higgs boson, the most important production channel for the
radion is $gg$ fusion, which has the lowest order in couplings.
In addition, $gg\to\phi$ gets further enhancement from the anomaly.
The production cross section at a hadronic collider with a center-of-mass
energy $\sqrt{s}$ is given by
\begin{equation}
\sigma(s) = \int^1_{m_\phi^2/s}\; \frac{dx}{x}\; g(x)\, 
g \left(\frac{m_\phi^2}{sx} \right) \; 
\frac{\alpha_s^2}{256 \pi \Lambda_\phi^2}\; \frac{m_\phi^2}{s}\;
\left| b_{\rm QCD} + y_t (1+ (1-y_t) f(y_t) \right |^2 \;,
\end{equation}
where $g(x)$ is the gluon parton distribution function 
at momentum fraction $x$.
As shown in Fig. \ref{pp-fig}, $gg\to \phi$ has substantially larger cross
sections than any other channels for all energies and masses.   We shall 
emphasize using this channel to detect the radion in the next section.
We have used the parameterization of CTEQ5L \cite{cteq5} for parton 
distribution functions. 

The associated Higgs production with a vector boson, $W$ or $Z$, is the golden
channel to search for the light Higgs boson at a relatively low energy machine,
such as the Tevatron.  This is because 
the $W$ or $Z$ can be tagged to reduce the huge QCD background and Higgs decays
mainly into the heaviest flavor $b\bar b$.  However, for 
the radion it is not the case because the light radion decays dominantly
into $gg$ so that heavy flavor cannot be tagged, but still the associated
$W$ or $Z$ boson can be tagged to reduce backgrounds.

The subprocess cross sections for $q\bar q' \to W \phi$ and $q\bar q \to
Z\phi$ are given by
\begin{eqnarray}
\hat \sigma(q\bar q'\to W\phi) &=& 
\frac{g^2}{192 \pi \Lambda_\phi^2} \; 
\frac{m_W^4}{(\hat s-m_W^2)^2} 
\sqrt{1 + \frac{(m_W^2 - m_\phi^2)^2}{\hat s^2} - 
\frac{2(m_W^2 + m_\phi^2)}{\hat s} }
\;\nonumber \\
&\times &
 \frac{m_\phi^4 + m_W^4 +10 m_W^2 \hat s + \hat s^2 -2m_\phi^2(m_W^2+\hat s)}
{3 m_W^2 \hat s}
\;, \\
\hat \sigma(q\bar q\to Z\phi) &=& \frac{g^2 ( {g_L^q}^2 + {g_R^q}^2)}{96 \pi 
\cos^2 \theta_{\rm w} \Lambda_\phi^2} \; 
\frac{m_Z^4}{(\hat s-m_Z^2)^2} 
\sqrt{1 + \frac{(m_Z^2 - m_\phi^2)^2}{\hat s^2} - \frac{2(m_Z^2 + m_\phi^2)}
{\hat s} }
\; \nonumber \\
&\times &
 \frac{m_\phi^4 + m_Z^4 +10 m_Z^2 \hat s + \hat s^2 -2m_\phi^2(m_Z^2+\hat s)}
{3 m_Z^2 \hat s}
\;,
\end{eqnarray}
where $g^f_{L,R} = T_{3f} - Q_f \sin^2 \theta_{\rm w}$ are the chiral
$\bar f fZ$ couplings, and $\hat s$ is the square of the 
center-of-mass energy of the incoming partons.

The $WW$ and $ZZ$ fusion processes will give an increasing cross section with
energy.  They are useful production channels 
at the LHC, especially, when energetic and forward jets are tagged 
\cite{jet-tag} while the other channels would not give rise to energetic nor 
forward jets.  There are a number of subprocesses, e.g., $u d \to u d \phi$
($WW$ and $ZZ$ fusion), $u u \to u u \phi$ ($ZZ$ fusion only), 
$u s \to d c \phi$ ($WW$ fusion only),  etc.  Here we only present the formula
for the subprocess $u d \to u d \phi$.  The other subprocesses can be obtained
from this by removing the interference term and replacing with appropriate
chiral couplings. The spin- and color-averaged amplitude-squared for 
$u (p_1) d (p_2) \to d (q_1) u(q_2) \phi (k)$ is given by
\begin{eqnarray}
\overline{\sum} \left| {\cal M} \right |^2 &=& 
\frac{4 g^4 m_W^4}{\Lambda_\phi^2} 
\;\frac{p_1 \cdot p_2 \; q_1 \cdot q_2}{[(p_1-q_1)^2 - m_W^2]^2 \;
[(p_2-q_2)^2 - m_W^2]^2 }\;, \nonumber \\
&+& \frac{16 g^4 m_Z^4}{\cos^4 \theta_{\rm w} \Lambda_\phi^2}
\;\frac{( {g_L^d}^2 {g_R^u}^2 + {g_L^u}^2 {g_R^d}^2) \;
     p_1 \cdot q_1 \; p_2 \cdot q_2
 + ( {g_L^d}^2 {g_L^u}^2 + {g_R^d}^2 {g_R^u}^2 ) \; 
     p_1\cdot p_2 \; q_1 \cdot q_2 }
{[(p_1-q_2)^2 - m_Z^2]^2 \; [(p_2-q_1)^2 - m_Z^2]^2 }\; \nonumber \\
&+& 
\frac{16 g^4 m_W^2 m_Z^2}{3 \cos^2 \theta_{\rm w} \Lambda_\phi^2}
\;\frac{ g_L^{d} g_L^{u}\;  p_1 \cdot p_2 \; q_1 \cdot q_2 }
{ [(p_1-q_1)^2 - m_W^2]\;  [(p_2-q_2)^2 - m_W^2] 
  [(p_1-q_2)^2 - m_Z^2] \; [(p_2-q_1)^2 - m_Z^2] } \;.
\end{eqnarray}

The associated production with a $t\bar t$ pair is also nonnegligible because
of the large top-quark mass. 
The formulas for $q\bar q \to t\bar t \phi$ and $gg\to t\bar t \phi$ are more 
complicated and thus presented in the appendix.  

The production cross sections for these channels versus the center-of-mass
energy and versus $m_\phi$ are given in Fig. \ref{pp-fig}(a) and (b),
respectively.  It is clear that $\sigma(gg\to \phi)$ is at least two orders of
magnitude larger than the other channels.     It is only this channel that
the production of radion is substantially larger than the production of the
Higgs boson, in the case of $\Lambda_\phi=v$.  In the next section, we
shall study a few decay modes of the radion using this $gg\to \phi$ production
channel.

\subsection{$e^+ e^-$ Colliders}

At $e^+ e^-$ colliders, the radion is produced via
\begin{eqnarray}
e^+ e^- &\to& Z \phi \;\;\;\; \mbox{(Higgs bremstralung)} \nonumber \\
e^+ e^- &\to& \nu \bar \nu \phi \;\;\;\; \mbox{($WW$ fusion)} \nonumber \\
e^+ e^- &\to& e^+ e^- \phi \;\;\;\; \mbox{($ZZ$ fusion)} \nonumber \;.
\end{eqnarray}

The differential cross section for the process $e^+ e^- \to Z\phi$ 
at a center-of-mass energy $\sqrt{s}$ is given by
\begin{equation}
\frac{d\sigma}{d\cos\theta}=\frac{g^2 ({g_L^e}^2 + {g_R^e}^2)}{32 \pi 
\cos^2 \theta_{\rm w} \Lambda_\phi^2}  \;
\frac{m_Z^4}{(s-m_Z^2)^2} \left( 1+ \frac{(m_Z^2 -t)(m_Z^2 -u)}{s m_Z^2}\right)
\sqrt{1 + \frac{(m_Z^2 - m_\phi^2)^2}{s^2} - \frac{2(m_Z^2 + m_\phi^2)}{s} }
\;,
\end{equation}
where $t,u= -\frac{(s-m_Z^2 -m_\phi^2)}{2}(1 \mp \alpha \cos\theta)$,
and $\alpha=\sqrt{s^2+m_\phi^4+m_Z^4 -2m_Z^2 m_\phi^2 -2m_Z^2 s -2m_\phi^2 s}/
(s-m_Z^2 - m_\phi^2)$.  The angular
distribution can be easily integrated to obtain the total cross section
\begin{equation}
\sigma = \frac{g^2 ( {g_L^e}^2 + {g_R^e}^2)}{32 \pi 
\cos^2 \theta_{\rm w} \Lambda_\phi^2} \; 
\frac{m_Z^4}{(s-m_Z^2)^2} 
\sqrt{1 + \frac{(m_Z^2 - m_\phi^2)^2}{s^2} - \frac{2(m_Z^2 + m_\phi^2)}{s} }
\; \frac{m_\phi^4 + m_Z^4 +10 m_Z^2 s + s^2 -2m_\phi^2(m_Z^2+s)}{3 m_Z^2 s}
\;.
\end{equation}

There are also contributions from $WW$ and $ZZ$ fusion.
The amplitude-squared for $WW$ fusion: $e^-(p_1) e^+(p_2) \to \nu (q_1)
\bar \nu(q_2) \phi$, and for $ZZ$ fusion: $e^-(p_1) e^+(p_2) \to e^- (q_1)
e^+(q_2) \phi$, 
summed over final-state spins and averaged over initial spins are given by
\begin{eqnarray}
\overline{\sum} \left|M \right|^2_{\rm WW}
 &=& \frac{4 g^4 m_W^4}{\Lambda_\phi^2}
\;\frac{p_1 \cdot q_2 \; p_2 \cdot q_1}{[(p_1-q_1)^2 - m_W^2]^2 \;
[(p_2-q_2)^2 - m_W^2]^2 }\;, \\
\overline{\sum} \left|M \right|^2_{\rm ZZ} &=& \frac{16 g^4 m_Z^4}
{\cos^4 \theta_{\rm w} \Lambda_\phi^2}
\;\frac{( {g_L^e}^4 + {g_R^e}^4) p_1 \cdot q_2 \; p_2 \cdot q_1 +
2 {g_L^e}^2 {g_R^e}^2 \; p_1\cdot p_2 \; q_1 \cdot q_2 }
{[(p_1-q_1)^2 - m_Z^2]^2 \; [(p_2-q_2)^2 - m_Z^2]^2 }\;.
\end{eqnarray}

\begin{figure}[t]
\centering
\includegraphics[width=5.5in]{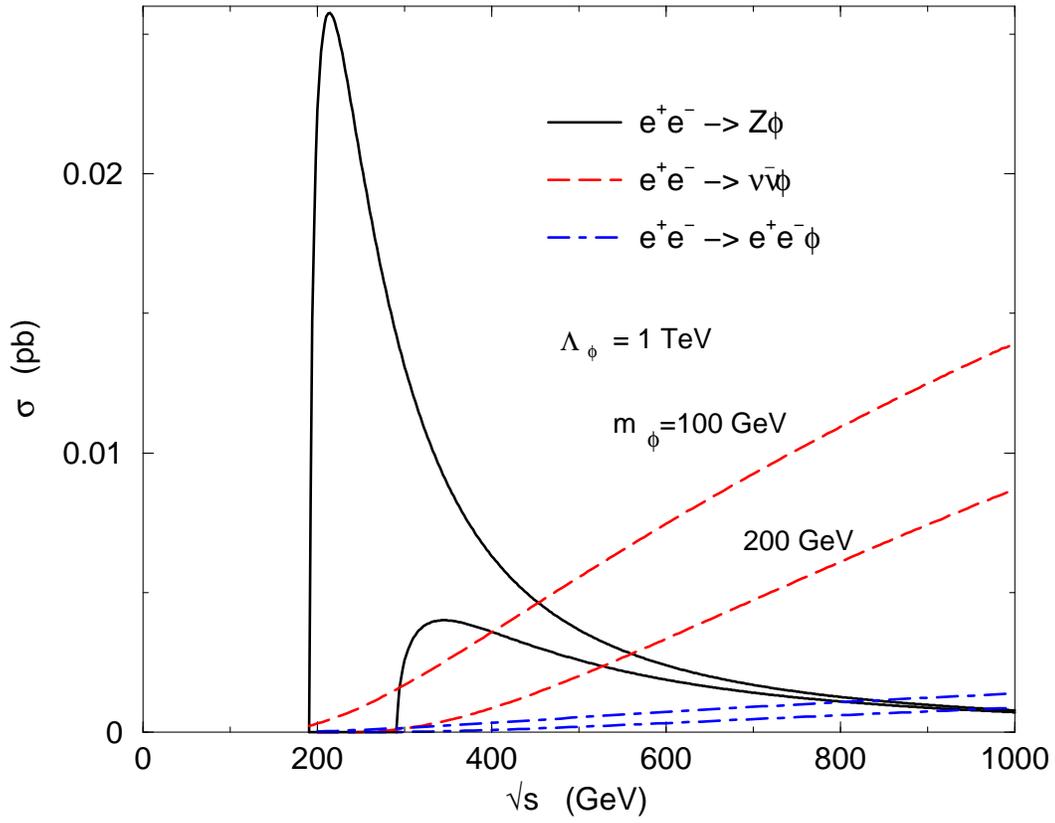}
\includegraphics[width=5.5in]{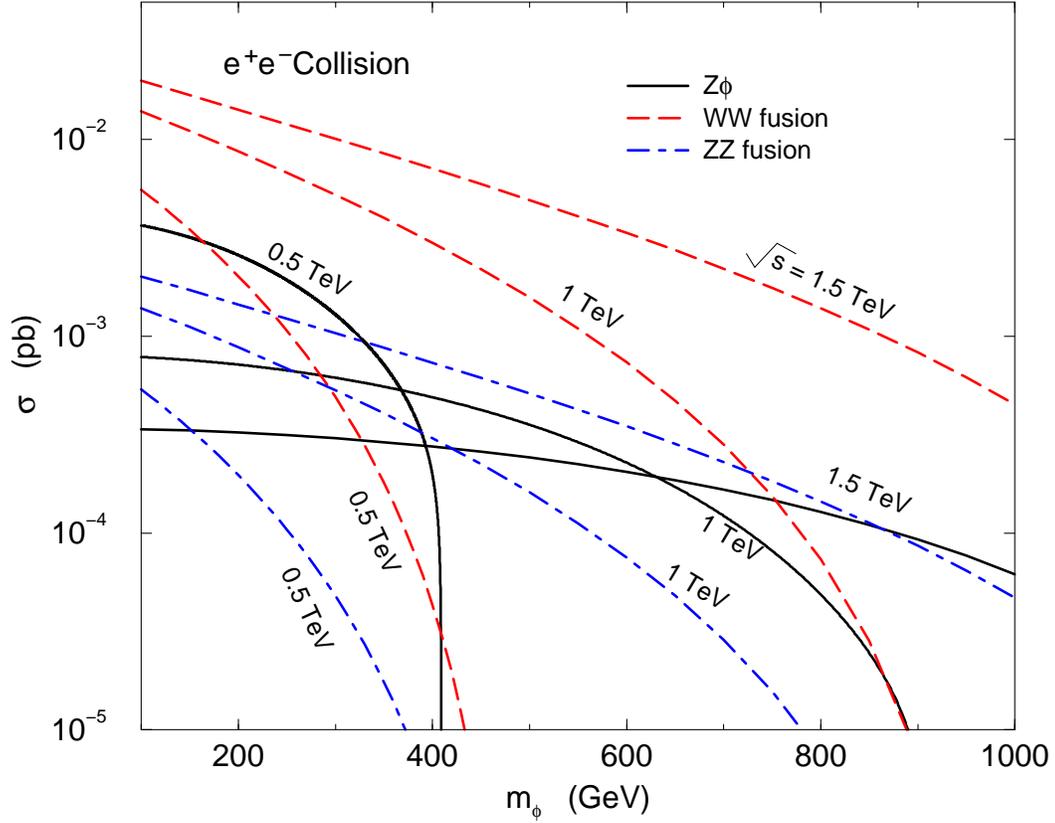}
\vskip0.1in
\caption{\label{ee-fig} \small
(a) Production cross sections versus the center-of-mass energy $\sqrt{s}$ for
$e^+ e^- \to Z\phi$, $e^+ e^- \to \nu \bar \nu \phi$ ($WW$ fusion), and 
$e^+ e^- \to e^+ e^- \phi$ ($ZZ$ fusion).
(b) Production cross sections versus $m_\phi$.
}
\end{figure}

The above three processes at $e^+ e^-$ colliders are not particularly 
different from the corresponding ones of the Higgs boson.  Given $\Lambda_\phi
=v$ they would be the same.  At any rates, the production of these
processes are given in Fig. \ref{ee-fig}(a) and (b) for cross sections 
versus $\sqrt{s}$ and $m_\phi$, respectively.

\subsection{$\gamma\gamma$ Colliders}

The radion can be directly produced in $\gamma \gamma$ collisions via the 
triangular loops  of $W$ boson and top quark, as well as the coupling from the
trace anomaly of the gauge group.  The production cross section at a 
center-of-mass energy $\sqrt{s_{\gamma\gamma}}$ of $\gamma\gamma$ collision 
is given by
\begin{equation}
\hat \sigma = \frac{\alpha_{\rm em}^2}{64 \pi \Lambda_\phi^2}\, 
\frac{s_{\gamma\gamma}}{m_\phi} \delta(\sqrt{s_{\gamma\gamma}} - m_\phi)\;
\left| b_2 + b_Y - (2+3y_W +3y_W (2-y_W)f(y_W) ) + \frac{8}{3}
y_t ( 1+ (1-y_t)f(y_t) )  \right |^2 \;.
\end{equation}

\begin{figure}[t]
\centering
\includegraphics[width=5.5in]{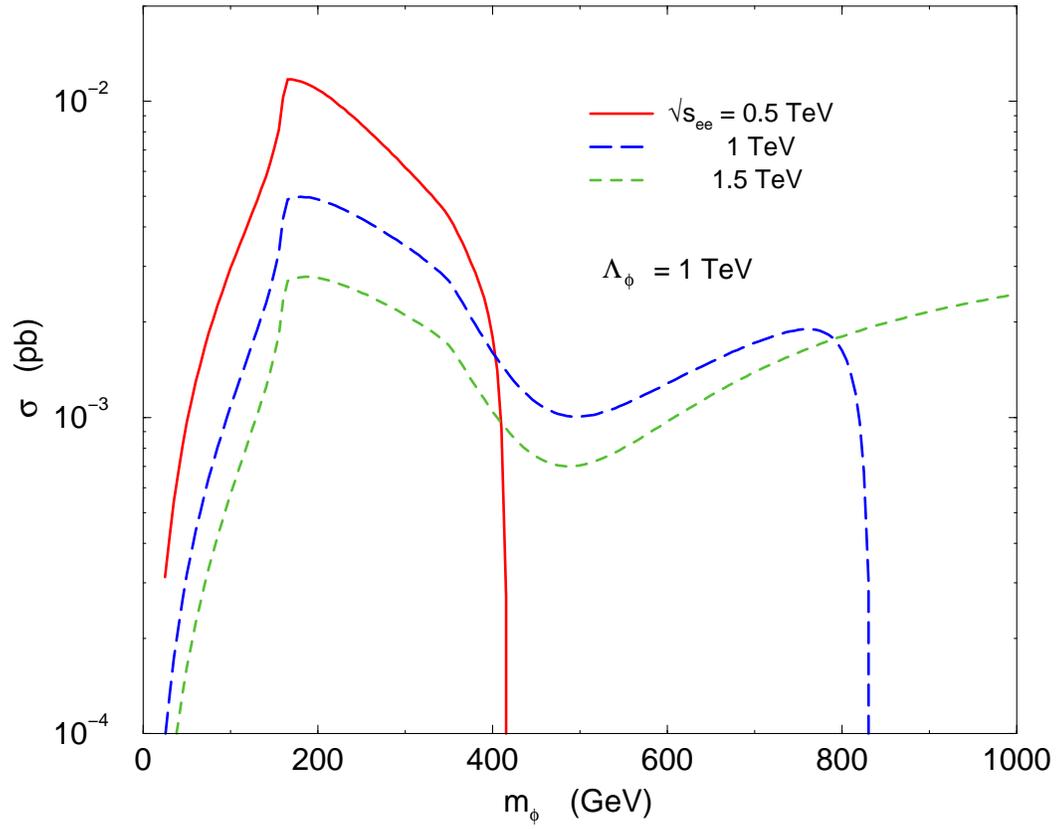}
\vskip0.1in
\caption{\label{aa-fig} \small
Production cross sections for the radion versus $m_\phi$ at linear
$e^+ e^-$ colliders in the backscattering $\gamma\gamma$ mode.
}
\end{figure}

Nearly monochromatic photon collider is achievable with the technique
of laser backscattering \cite{telnov} on $e^+ e^-$ colliders.  The 
unpolarized photon flux resulting from the laser backscattering is 
\begin{equation}
f(x) = \frac{1}{D_\xi} \left(1 - x + \frac{1}{1-x} - \frac{4x}{\xi(1-x)}
             + \frac{4 x^2}{\xi^2 (1-x)^2} \right ) \;,
\end{equation}
where $D_\xi = (1 - 4/\xi - 8/\xi^2 ) \log(1+\xi) + 1/2 
+ 8/\xi - 1/2(1+\xi)^2$ and $x$ is momentum fraction of the energy of the
incoming electron carried away by the backscattered photon.
The choice of $\xi=4.8$ optimizes the monochromaticity of the photon beam.
The production cross sections for radion versus the radion mass for a few
linear collider energies are given in Fig. \ref{aa-fig}.

\section{Experimental Detection}

\subsection{Hadronic Collisions}

\underline{$gg$}.
The most obvious difference in decay modes between the radion and the SM
Higgs boson is the $\phi\to gg$ mode at the low radion mass ($\alt 150$ GeV).  
Other than that the decay modes of the radion with moderate to heavy mass 
are similar to those of the SM Higgs boson.  Here we study the detection of
radion using $gg, b\bar b, WW$, and $ZZ$ modes.

We first look at the sensitivity to the past and the present experimental data.
The UA2 collaboration had searched for particles that decay into a pair of 
jets in $p\bar p$ collisions at $\sqrt{s}=630$ GeV \cite{ua2}.  Due to low
collision energy, they only searched particles of mass between about 100
and 300 GeV.  In their searches, they particularly searched for extra
$W$ and $Z$ bosons, but the sensitivity for other particles that decay into 
a pair of jets is similar.   On the other hand, the data from CDF \cite{cdf} 
give limits on a heavier mass 
range from 200 GeV to about 1 TeV.  They can be combined in the sense that
the whole mass range from 100 GeV to 1 TeV is covered.

\begin{figure}[t]
\centering
\includegraphics[width=5.5in]{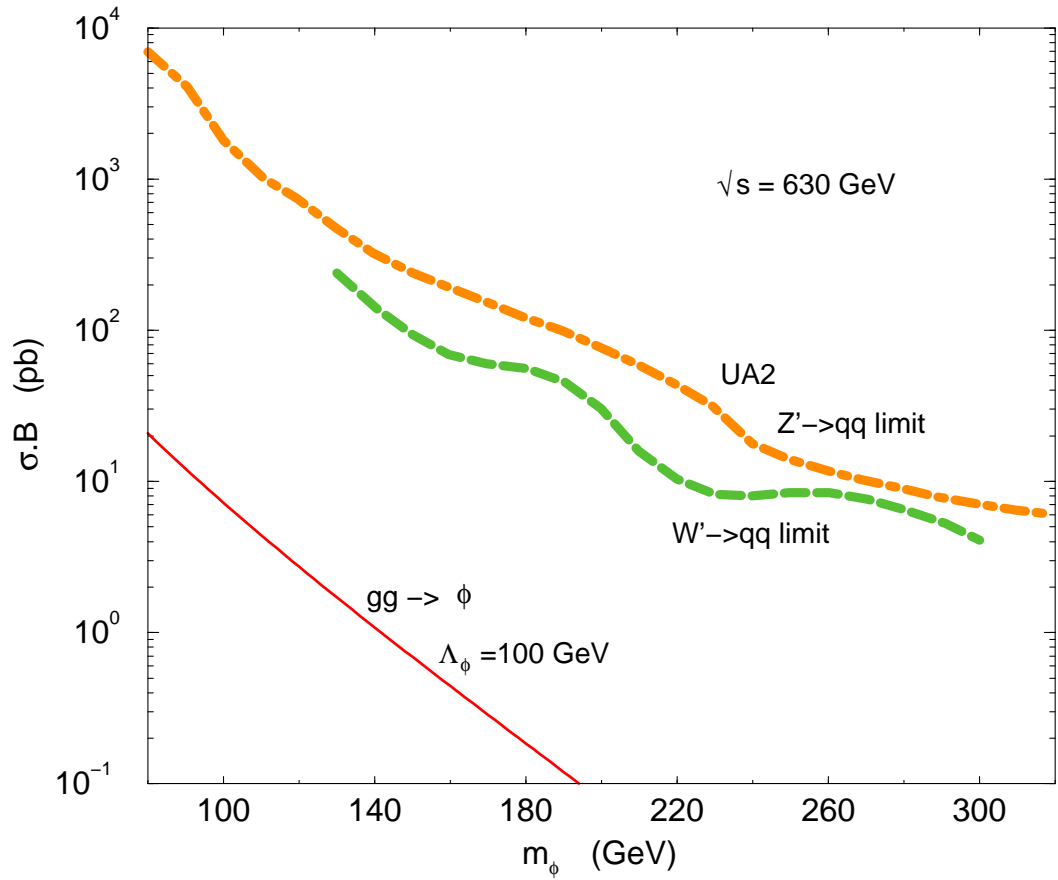}
\vskip0.1in
\caption{\label{ua2-fig} \small
Dijet production cross section for the radion using $gg\to \phi$ channel at
$\sqrt{s}=630$ GeV, assuming $B(\phi \to gg)=1$.  The UA2 90\% C.L. upper 
limits on dijet production from decays of heavy bosons ($Z'$ and $W'$) are
shown.
}
\end{figure}

The UA2 limits on their searches for $Z'$ and $W'$ are shown in 
Fig. \ref{ua2-fig}.  We used a thick line to show the fact that the
UA2 studied a number of $W'$ and $Z'$ models and the curves are closely
packed together.  In Fig. \ref{ua2-fig}, we have used a very small value for 
$\Lambda_\phi=100$ GeV, in order to make the radion curve appear on the 
figure.  However, radion production is far below the 90\% C.L. upper 
limits offered by UA2, even for such a small $\Lambda_\phi$.

\begin{figure}[t]
\centering
\includegraphics[width=5.5in]{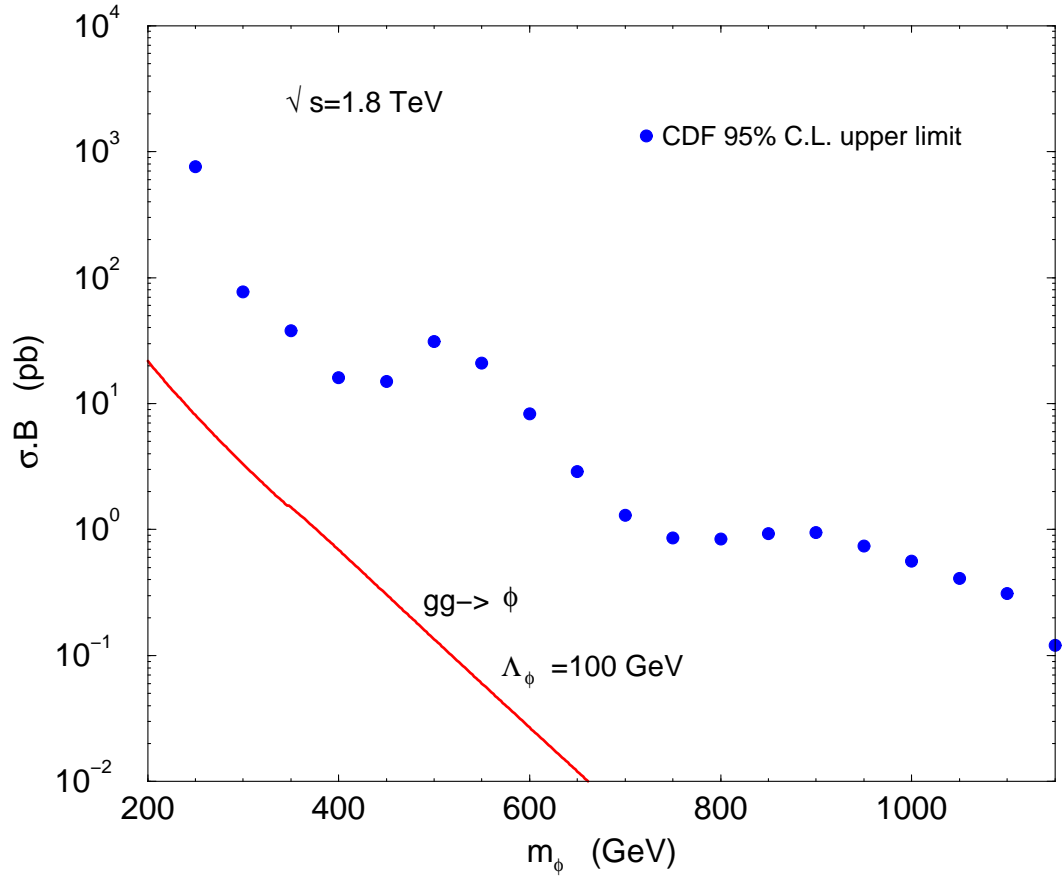}
\vskip0.1in
\caption{\label{cdf-fig} \small
Dijet production cross section for the radion using $gg\to \phi$ channel at
$\sqrt{s}=1.8$ TeV, assuming $B(\phi \to gg)=1$.  The CDF 95\% C.L. upper 
limit on dijet production from the decay of a new particle is
shown.
}
\end{figure}

Similarly, in Fig. \ref{cdf-fig} we show the 
dijet production of the radion for mass above 200 GeV, assuming the radion 
still decays dominantly into $gg$ (though it is not the case as shown in 
Fig. \ref{decay}.)
\footnote{In Ref. \cite{GRW}, it is shown that in some cases the mixing of 
the radion and the Higgs boson causes the radion to decay dominantly into
$gg$ even for moderate to heavy radion mass.}
But still the radion production is substantially below the 95\% C.L. upper
limit offered by CDF, even though we used a $\Lambda_\phi=100$ GeV.
The conclusion here is that the present dijet background is far too large for 
the radion in the dijet mode to be sensitive.

\begin{figure}[t]
\centering
\includegraphics[width=5.5in]{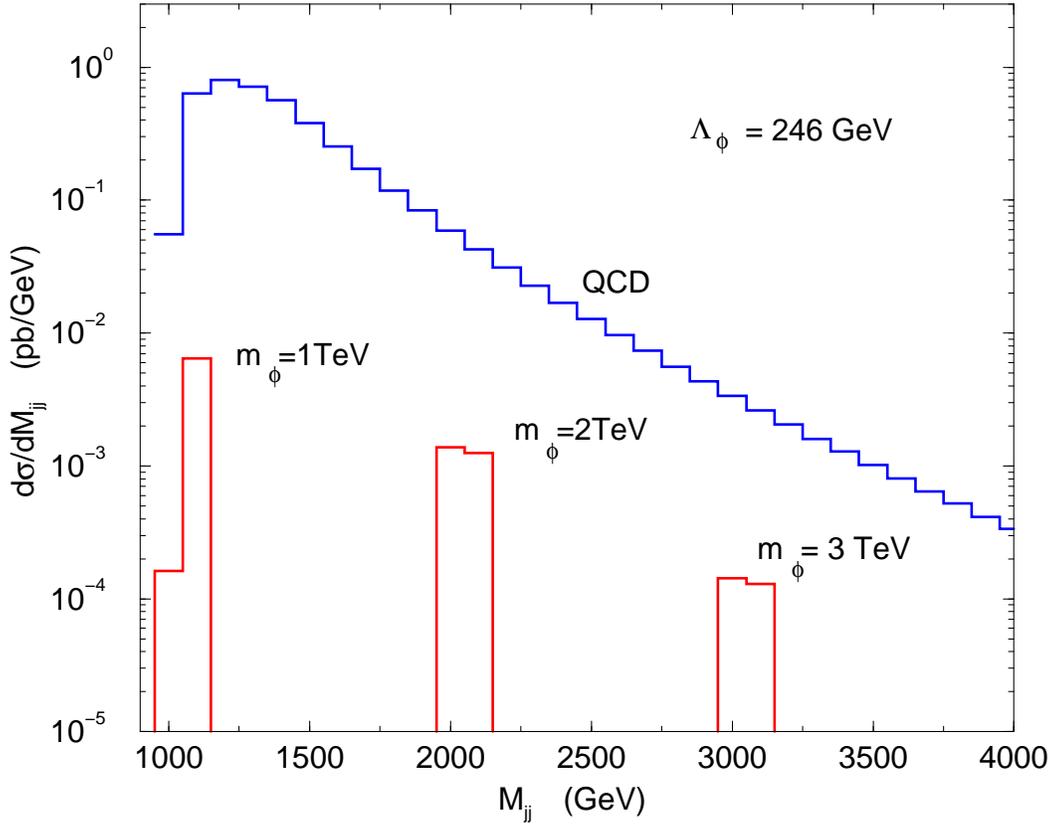}
\vskip0.1in
\caption{\label{lhc-dijet} \small
Differential cross section versus dijet invariant mass for the QCD background
and for radion signals (1--3 TeV) using $gg\to \phi \to gg$ at the LHC.
We have assumed $B(\phi \to gg)=1$.  
We have imposed a strong $p_T(j) >500$ GeV on
each jet, $|\cos\theta^*|<2/3$ and $|y_j|<2$, 
and employed a smearing of $\Delta E/E = 100\%/\sqrt{E}$ on the jets.  
}
\end{figure}

We also investigate the possibility of dijet mode at the LHC.  We emphasize
the heavy radion with $m_\phi > 1$ TeV and naturally with $\Lambda_\phi 
\agt 246$ GeV.  We show in Fig. ~\ref{lhc-dijet} the differential cross section
versus the dijet invariant mass for the QCD background (we have employed the
leading-order calculation) and for the radion signal
$gg\to \phi \to gg$.  Here we have also assumed $B(\phi \to gg)=1$ to 
illustrate the best scenario.  We have imposed a strong $p_T(j) >500$ GeV on
each jet, $|\cos\theta^*|<2/3$, and $|y_j|<2$ to suppress the QCD background,
and employed a smearing of $\Delta E/E = 100\%/\sqrt{E}$ on the jets.  We
used a bin resolution of 100 GeV and the signals essentially spread into
2 bins (where we have assumed the intrinsic width of the radion is negligible.)
In Fig. \ref{lhc-dijet}, we see that the signal is about 1--2\% of the 
QCD background in the corresponding bins.  Therefore, it is very
difficult to identify the signal
given such large systematic uncertainties in dijet production.
Should we have used a much smaller $\Lambda_\phi$ the signal would have 
been large enough to be identified.  However, such a small $\Lambda_\phi$
but with a TeV radion mass is rather unnatural.

\begin{figure}[t]
\centering
\includegraphics[width=5.5in]{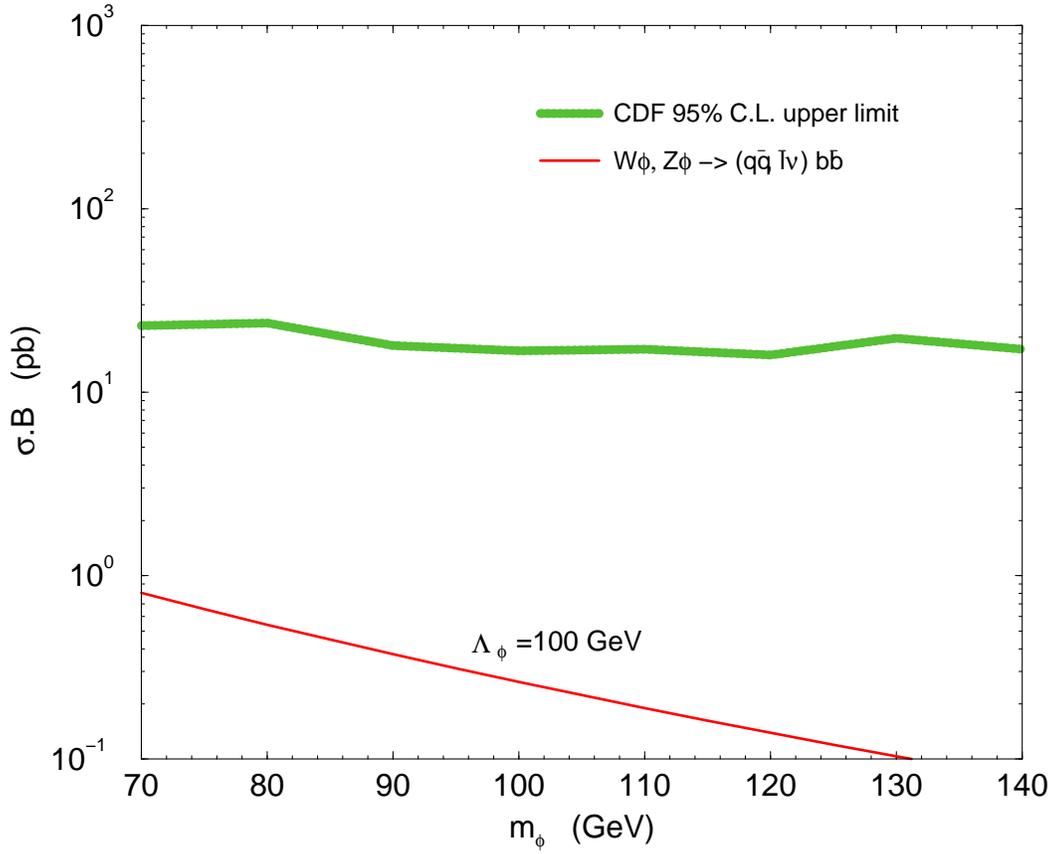}
\vskip0.1in
\caption{\label{cdf-vh-fig} \small
Cross section for the radion production using $p\bar p\to V\phi \to
(q\bar q + l\bar \nu) b \bar b$ $(V=W,Z)$ at $\sqrt{s}=1.8$ TeV, 
assuming $B(\phi \to b\bar b)=0.1$.  
The CDF 95\% C.L. upper limit on $VH^0 \to (q\bar q + l\bar \nu) b \bar b$
is shown.
}
\end{figure}

\underline{$b\bar b$}.
Another possibility is using the $b\bar b$ mode of the radion, though
the branching ratio is only of order $0.1$ whereas the $b\bar b$ mode of 
the SM Higgs boson is close to 1.  CDF collaboration \cite{cdf-wh} has 
searched for the $WH$ and $ZH$ production and obtained 95\% C.L. upper 
limits on their production.  We show the production of $W\phi$ and $Z\phi$
with $\phi \to b\bar b$ and $W$ and $Z$ decay leptonically and hadronically
in Fig. \ref{cdf-vh-fig}, where the 95\% C.L. upper limit obtained by CDF
is also shown.  It is clear that only the case with $\Lambda_\phi$ much
less 100 GeV would it be sensitive to the data.
As a consequence, the detection of the radion seems much worse than the
SM Higgs boson because of depletion of $b\bar b$ mode.

\begin{figure}[t]
\centering
\includegraphics[width=5.5in]{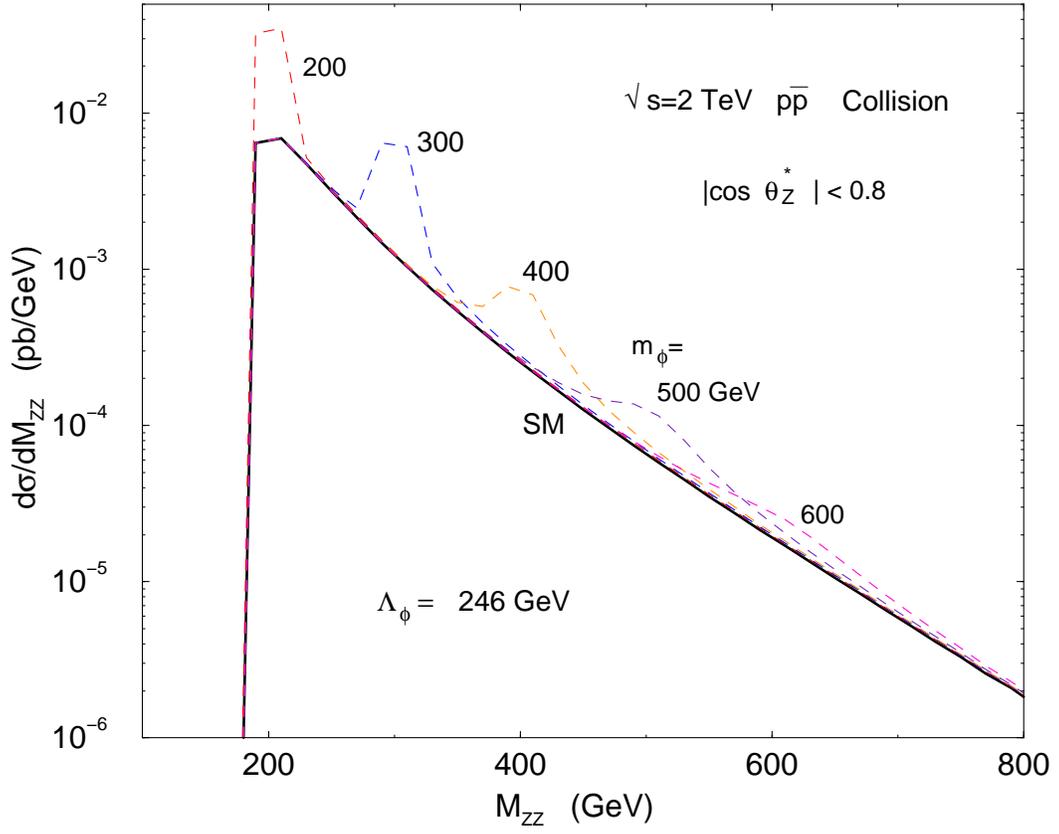}
\vskip0.1in
\caption{\label{phi-zz-teva} \small
Invariant mass distribution $d\sigma/dM_{ZZ}$ for the radion signal
with $m_\phi=200-600$ GeV and the SM background $q\bar q \to ZZ$ at 
$\sqrt{s}=2$TeV $p\bar p$ collision, for $\Lambda_\phi=246$ GeV. A cut
of $|\cos\theta^*_Z|<0.8$ is imposed.
}
\end{figure}

\underline{$ZZ$}.
Since radion production by $gg$ fusion is substantially larger than the
SM Higgs boson, we shall see that using $WW$ and $ZZ$ modes the 
discovery of the radion is much better than the Higgs boson.
For radion heavier
than 180 GeV, $\phi \to ZZ \to 4l$ is the gold-plated mode.  The 
production is mainly via gluon fusion (see Fig. \ref{pp-fig}), and the final
state would be four charged leptons.  
We expect this mode remains the gold-plated one for the radion with masses
larger than about 180 GeV, 
as long as the $\Lambda_\phi$ and $m_\phi$ are not getting too large.
For illustration we show the invariant mass $M_{ZZ}$ distribution 
at the 2 TeV Tevatron in Fig. \ref{phi-zz-teva} for radion signals with 
$m_\phi=200-600$ GeV and $\Lambda_\phi=246$ GeV, together with the background 
from $q\bar q \to ZZ$ added.  The choice of $\Lambda_\phi=246$ GeV makes
the couplings of the radion the same as the SM Higgs boson, except for the
$gg$ and $\gamma\gamma$ couplings.
We have used an angular cut of $|\cos\theta^*_Z|<0.8$ on the 
center-of-mass scattering angle of the $Z$ boson pair.  From the figure
we can see peak structures for $m_\phi=200-400$ GeV, but only a small
bump for $m_\phi>400$ GeV, and almost no structure for $m_\phi>600$ GeV.
To account for observation we have to take into account the event rate.
The four-lepton decay mode of $ZZ$ has a branching ratio of 0.0045.
The event rate for $m_\phi=400$ GeV ($\sigma=0.032$ pb) is about 4.4 for
an integrated luminosity of 30 fb$^{-1}$ (Run IIb) and 
thus is marginal for discovery.
Therefore, any radion heavier than 400 GeV is hard to be detected in Run II.
In other words, using $gg\to \phi \to ZZ \to 4 l$ the radion with mass upto
about 400 GeV would be discovered at RunIIb of the Tevatron.

\begin{figure}[t]
\centering
\includegraphics[width=5.2in]{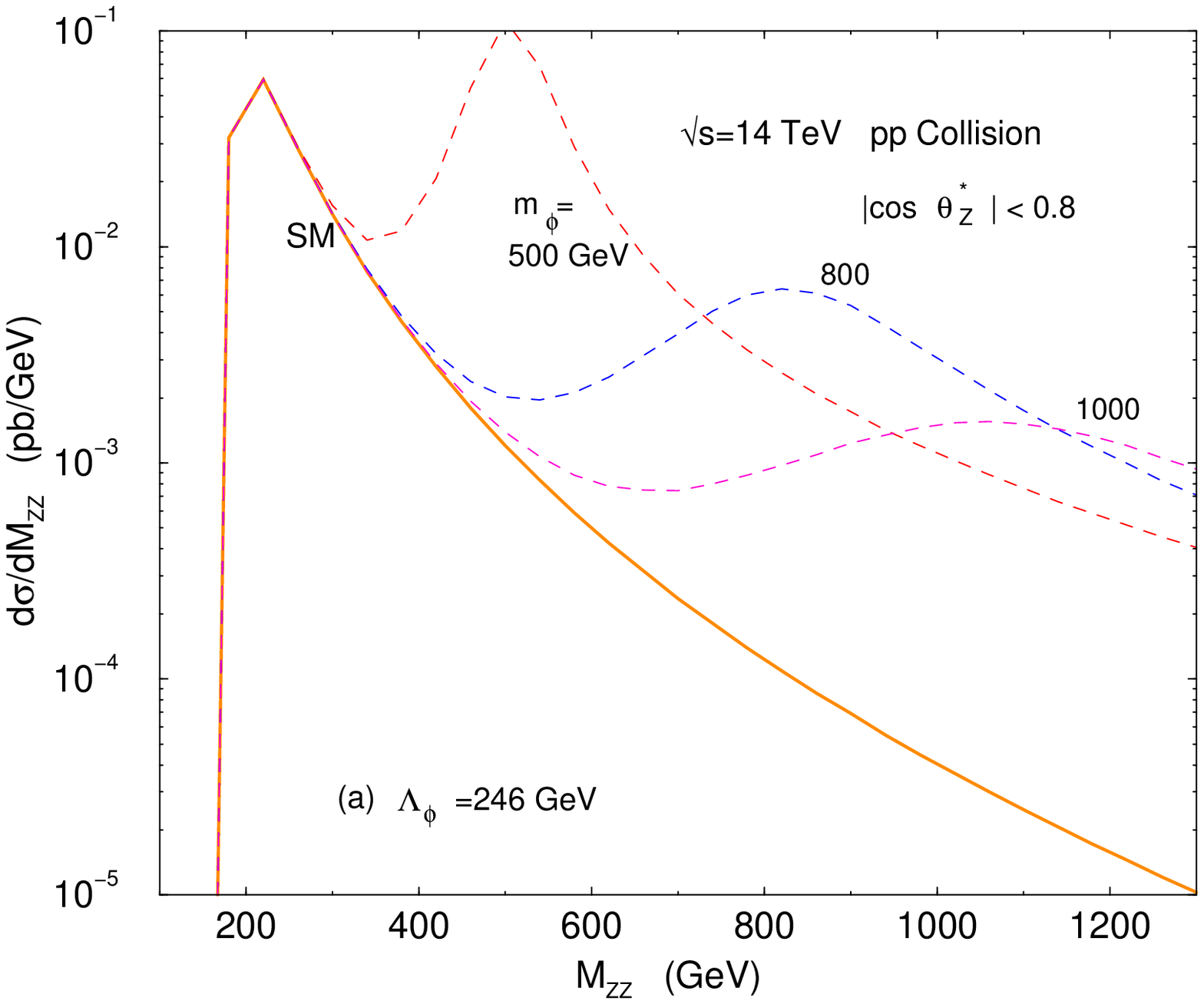}
\includegraphics[width=5.2in]{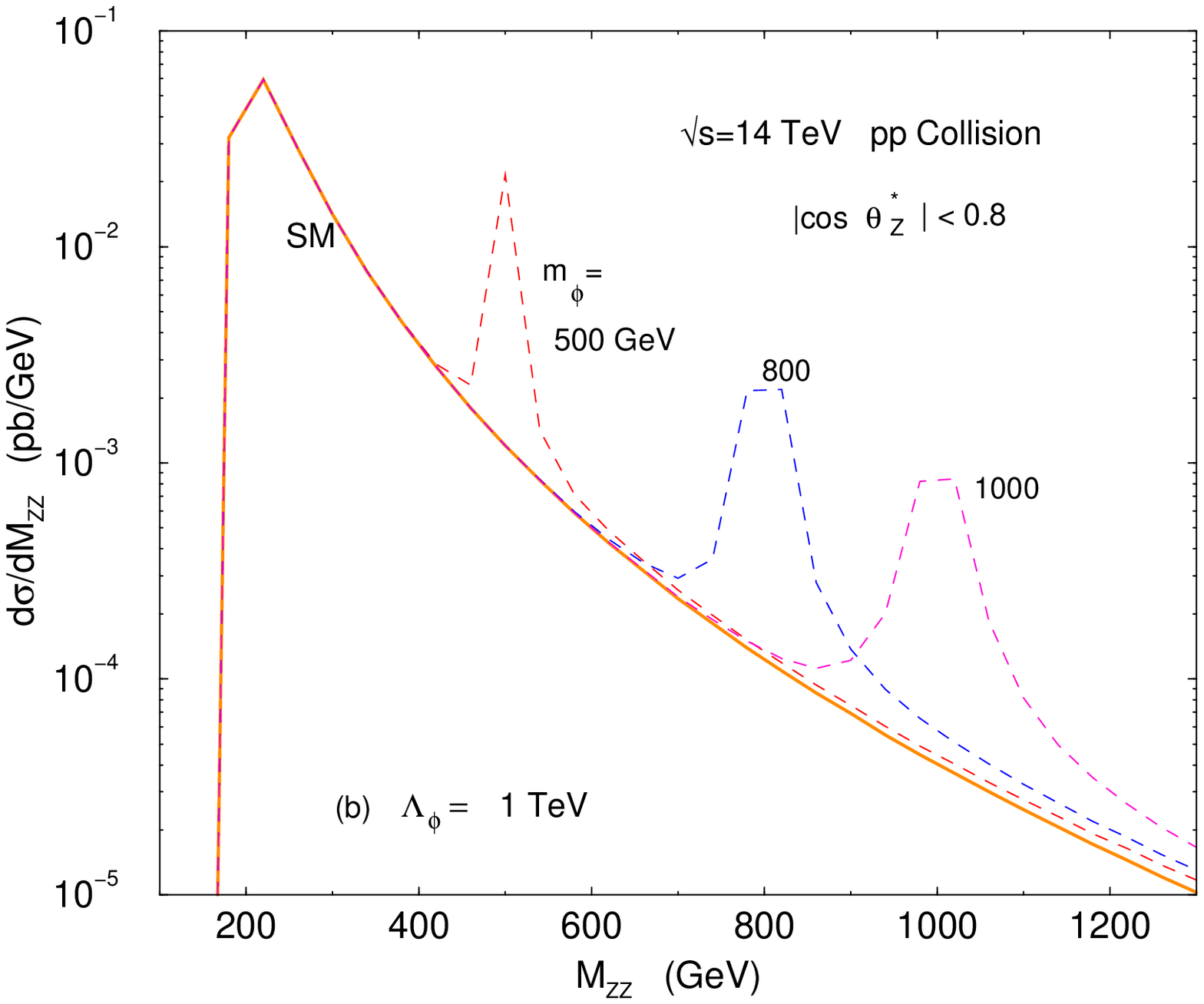}
\vskip0.1in
\caption{\label{phi-zz-lhc} \small
Invariant mass distribution $d\sigma/dM_{ZZ}$ for the radion signal
with $m_\phi=500,800, 1000$ GeV and the SM background $q\bar q \to ZZ$ at 
the LHC, for (a) $\Lambda_\phi=246$ GeV and (b) $\Lambda_\phi=1$ TeV.
A cut of $|\cos\theta^*_Z|<0.8$ is imposed.}
\end{figure}

The situation at the LHC will improve significantly.  We show in 
Fig. \ref{phi-zz-lhc} the invariant mass $M_{ZZ}$ distribution at the LHC
for $\Lambda_\phi=246$ GeV and 1 TeV.  We can see that the signal is way
above the background.  The detectability depends on event rates and 
clarity of a peak structure.
Since the width of the radion scales
as $1/\Lambda_\phi^2$, the peak structure starts to lose when $m_\phi>800$
GeV for the case of $\Lambda_\phi=246$ GeV.  For $\Lambda_\phi=1$ TeV, the
peak structure of the radion remains even for a 1 TeV radion.
The event rate for the 1 TeV radion to decay into
 four charged lepton is about 39
for $\Lambda_\phi=1$ TeV, which is more than enough for discovery 
given such a small background.  When $\Lambda_\phi$ becomes small we have to
take into account the unitarity of the calculation, but we do not concern
ourselves with it here.  For $\Lambda_\phi=1$ TeV unitarity should be safe.

\underline{$WW$}.
A recent study in Ref. \cite{hww} 
showed that the production and decay of the SM Higgs boson in
$gg\to H\to W^*W^* \to l \bar\nu \bar l \nu$ allows the detection of the
Higgs for $145 \alt m_H \alt 180$ GeV at a $3\sigma$ level.  For the case of 
radion, since the production cross section of $gg\to \phi$ is 
substantially larger, we expect the detectability of the radion
to be significantly better than the Higgs boson.
Based on the analysis performed in Ref. \cite{hww} we estimate the
signal and signal-to-background significance of the radion.
The results are shown in Table \ref{table1}.  We have used their background
numbers and the efficiencies, but with our own cross sections and branching 
ratios of the radion.  We show the results for two choices of $\Lambda_\phi=
0.246$ and 1 TeV.  Thus, we can see that if $\Lambda_\phi\approx v$ the radion
can be easily identified in the mass range of 140--190 GeV using the $W^*W^*$ 
decay mode.  Even if $\Lambda_\phi$ becomes as large as 1 TeV the radion
detectability is still better than the SM Higgs boson \cite{hww}.

\begin{table}[th]
\caption[]{\small \label{table1}
Cross sections for the radion signal in the channel
$gg\to \phi \to W^* W^* \to l \bar\nu \bar l \nu$ and the 
corresponding significance.  
The background cross sections are from Ref. \cite{hww}.  
}
\medskip
\begin{tabular}{c|cccccc}
\hline
$m_\phi$ (GeV) & 140  &  150  &  160  &  170  &  180  &  190  \\
\hline
\multicolumn{7}{c}{$\Lambda_\phi=246$ GeV} \\
\hline
$\sigma$ (fb) & 18   & 28   & 32   & 31  & 28  & 25 \\
bkgd (fb)     & 44   & 30   & 4.4  & 2.4 & 3.8 & 7.5 \\
$S/\sqrt{B}$ (30 fb$^{-1}$) & 15 & 28 & 83 & 108 & 78 & 18 \\
\hline
\multicolumn{7}{c}{$\Lambda_\phi=1$ TeV} \\
\hline
$\sigma$ (fb) & 1.1  & 1.7  & 1.9  & 1.8 & 1.7  & 1.5 \\
bkgd (fb)     & 44   & 30   & 4.4  & 2.4 & 3.8 & 7.5 \\
$S/\sqrt{B}$ (30 fb$^{-1}$) & 0.9 & 1.7 & 5.1 & 6.5 & 4.8 & 3.0 \\
\hline
\end{tabular}
\end{table}

\subsection{$e^+ e^-$ Colliders}

The search strategy for the radion at $e^+ e^-$ colliders is very similar to
that of Higgs bosons, because the production mechanisms are the same and the
decays are similar.  The only difference is that for the mass range
relevant to LEP search, the dominant decay mode of the radion is 
$\phi \to gg$ instead of $b\bar b$.  Thus, except for heavy flavor 
tagging, the radion or the Higgs decays into a pair of jets.  
With $b$-tagging the effectiveness against backgrounds for finding the Higgs 
boson is better than the radion.  Therefore, the limit on the radion would 
be lower than the limit on the Higgs boson, if not slightly lower,
provided that $\Lambda_\phi=v$.  If $\Lambda_\phi>v$ the limit 
would be worse, and vice versa.
The most updated LEP limit on the SM Higgs boson mass is 107.9 GeV 
\cite{lep-mh} using all decay channels of $Z$ and $H$.    

At higher $\sqrt{s}$, e.g., the next linear collider (NLC) energies, the
useful channels include also the $WW$ and $ZZ$ fusion.  These production
cross sections rise with $\sqrt{s}$ while the $H$-bremstralung mechanism
decreases with $\sqrt{s}$.  

So far, we have only studied cases with $\Lambda_\phi \le 1$ TeV.  What 
will happen to the signal when $\Lambda_\phi$ becomes as large as 
2 -- 3 TeV?  The obvious
answer is that various signals get smaller because the signal cross sections
scale by $1/\Lambda_\phi^2$.  

For those dijet channels ($jj$ and $b\bar b$) the already huge background
will further bury the even smaller signal.  For the $ZZ$ channel we have shown
that Run II with 30 fb$^{-1}$ integrated luminosity can marginally detect
a radion of mass $m_\phi=400$ GeV and $\Lambda_\phi=246$ GeV.  Therefore, 
at the Tevatron there is no hope of observing  the radion if $\Lambda_\phi
\ge 1$ TeV.
At the LHC the $ZZ\to 4\ell$ remains  the golden channel for larger
$\Lambda_\phi$  as long as there are enough signal events.   Take $m_\phi=1$
TeV the background is only about 1.8 events for a mass bin of 100 GeV around
1 TeV.  The signal goes down by $1/\Lambda_\phi^2$.  Therefore, for
$\Lambda_\phi=3$ TeV the signal event rate goes down to about 4, which may
only be marginally enough for discovery. 
The situation for the $WW$ channel would be similar, though it is not as 
good as the $ZZ$ channel.

At $e^+ e^-$ collider, again the signal cross section scales by
$1/\Lambda_\phi^2$.  The chance of discovering the radion becomes weaker
and weaker when $\Lambda_\phi$ gets larger.

\section{Conclusions}

In this work, we have studied the effective interactions, decay, production,
and detection of the radion, which is perhaps lighter than the Kaluza-Klein
excitations of gravitons in the Randall-Sundrum scenario with the bulk
modulus field being stabilized by the Goldberger-Wise mechanism.

The effective interactions of the radion are very similar to the SM Higgs
boson, except for the $\phi g g$ and $\phi\gamma\gamma$ couplings, which 
include the anomaly contribution.  Because of the anomalous $\phi gg$ 
coupling, the dominant
decay mode for $m_\phi$ lighter than about 150 GeV is the $gg$ mode, instead
of the $b\bar b$ mode.  However, once $m_\phi$ heavier than $WW$ and $ZZ$ 
thresholds the $WW$ and $ZZ$ modes dominate.

Again, the hadronic production of the radion is dominated by $gg\to \phi$.
The other production channels: $WW$ and $ZZ$ fusion, $W\phi, Z\phi$, and 
$t\bar t \phi$ at hadron colliders 
are just similar to those of the SM Higgs boson. 
At $e^+ e^-$ colliders, the production channels are similar to those of the
Higgs bosons.  

Although the radion decay is dominated by the $gg$ mode for $m_\phi \alt 150$
GeV and the production via $gg \to \phi$ is anomalously large, this 
signal $gg\to\phi \to gg$ is still buried under the dijet background, as 
demonstrated in Fig. \ref{ua2-fig} at $\sqrt{s}=630$ GeV and in 
Fig. \ref{cdf-fig} at $\sqrt{s}=1.8$ TeV.  Similar situation would hold at
the LHC. We have also shown that 
$q \bar q' \to V\phi \to (l\nu + q q) gg$ is not useful. 

On the other hand, we have studied the $ZZ$ and $WW$ decay modes of the radion
with the radion produced by $gg\to \phi$.  Based on a study on 
$H\to W^* W^* \to l \bar \nu \bar l \nu$
we have shown that the detection of radion is much better than the Higgs boson,
because of the 
anomalously large production of $gg\to \phi$ (see Table \ref{table1}).
The gold-plated mode $\phi\to ZZ\to
l\bar l l \bar l$ is the cleanest signature for radion for $m_\phi \agt 180$
GeV.  At Run II with a luminosity of 30 fb$^{-1}$ up to $m_\phi=400$ GeV
with $\Lambda_\phi=v$ can be discovered.  The situation at the LHC improves
significantly.  Even with $\Lambda_\phi=1$ TeV, the radion with mass all the 
way to 1 TeV can be discovered with a sharp peak.  If $\Lambda_\phi$ is 
relatively small, the peak structure becomes a broad bump, but is still much
larger than the SM background.  We, therefore, urge the experimenters to
search for the radion using the $ZZ$ and $WW$ modes.

\section*{\bf Acknowledgments}
This research was supported in part by the U.S.~Department of Energy under
Grants No. DE-FG03-91ER40674 and by the Davis Institute for High Energy 
Physics, and in part by a grant from the NSC of Taiwan R.O.C.

\appendix
\section{}
\subsection{$q\bar q\;, gg \to t\bar t\phi$}

Here we present the helicity amplitudes for $q (p_1,j)\; \bar q(p_2,i) \to
t(q_1,k) \;\bar t(q_2,l) \; \phi(k_1)$, where the momenta and color indices 
are denoted in the parentheses.  
\begin{eqnarray}
i{\cal M}_1 &=& -i \frac{g_s^2 m_t}{\Lambda_\phi}\, T^a_{ij} T^a_{kl}
\frac{1}{(p_1+p_2)^2}\; \frac{1}{(q_1+k_1)^2 - m_t^2}\;
\bar v(p_2) \gamma^\mu u(p_1)\; \bar u(q_1)\left( \overlay{/}{q}_1 +
\overlay{/}{k}_1 + m_t \right) \gamma_\mu v(q_2)\;, \\ 
i{\cal M}_2 &=& -i \frac{g_s^2 m_t}{\Lambda_\phi}\,
T^a_{ij} T^a_{kl}\frac{1}{(p_1+p_2)^2 }\;\frac{1}{(q_2+k_1)^2 - m_t^2}\; 
\bar v(p_2) \gamma^\mu u(p_1)\; \bar u(q_1) \gamma_\mu \left( 
- \overlay{/}{k}_1 - \overlay{/}{q}_2 + m_t \right) v(q_2)\;.
\end{eqnarray}

There are eight Feynman diagrams contributing to the subprocess
$g (p_1,a)\; g (p_2,b) \to t (q_1,i)\; \bar t (q_2,j)\; \phi(k_1)$.  
They are given by
\begin{eqnarray}
i{\cal M}_1 &=& i g_s^2 (T^a T^b)_{ij}\frac{m_t}{\Lambda_\phi}
\frac{1}{(q_1+k_1)^2 - m_t^2}\;
\frac{1}{(p_2-q_2)^2 - m_t^2}\;
\bar u(q_1) \left( \overlay{/}{q}_1 + \overlay{/}{k}_1 + m_t \right) 
\overlay{/}{\epsilon}(p_1) \left( \overlay{/}{p}_2 - \overlay{/}{q}_2 + 
m_t \right) \overlay{/}{\epsilon}(p_2) v(q_2)\;, \\
i{\cal M}_2 &=& i g_s^2 (T^a T^b)_{ij}\frac{m_t}{\Lambda_\phi}
\frac{1}{(q_1-p_1)^2 - m_t^2}\;
\frac{1}{(p_2-q_2)^2 - m_t^2}\;
\bar u(q_1) \overlay{/}{\epsilon}(p_1) \left( \overlay{/}{q}_1- 
\overlay{/}{p}_1 + m_t \right) 
 \left( \overlay{/}{p}_2 - \overlay{/}{q}_2 + m_t \right) 
\overlay{/}{\epsilon}(p_2) v(q_2)\;, \\
i{\cal M}_3 &=& i g_s^2 (T^a T^b)_{ij}\frac{m_t}{\Lambda_\phi}
\frac{1}{(q_1-p_1)^2 - m_t^2}\;
\frac{1}{(q_2+k_1)^2 - m_t^2}\;
\bar u(q_1) \overlay{/}{\epsilon}(p_1) \left( \overlay{/}{q}_1- 
\overlay{/}{p}_1 + m_t \right) \overlay{/}{\epsilon}(p_2) 
 \left(- \overlay{/}{q}_2 - \overlay{/}{k}_1 + m_t \right) v(q_2)\;, \\
i{\cal M}_4 &=& i g_s^2 (T^a T^b - T^b T^a)_{ij}\frac{m_t}{\Lambda_\phi}
\frac{1}{(p_1+p_2)^2 }\; \frac{1}{(q_1+k_1)^2 - m_t^2}\; 
\bar u(q_1) \left( \overlay{/}{q}_1+ \overlay{/}{k}_1 + m_t \right) 
 \gamma^\rho v(q_2) \; \nonumber \\
&&\times \left[ (p_1-p_2)_\rho \epsilon(p_1)\cdot \epsilon(p_2) +
2 p_2 \cdot \epsilon(p_1) \epsilon(p_2)_\rho -
2 p_1 \cdot \epsilon(p_2) \epsilon(p_1)_\rho \right ] \;, \\
i{\cal M}_5 &=& i g_s^2 (T^a T^b - T^b T^a)_{ij}\frac{m_t}{\Lambda_\phi}
\frac{1}{(p_1+p_2)^2 }\; \frac{1}{(q_2+k_1)^2 - m_t^2}\; 
\bar u(q_1) \gamma^\rho 
\left(- \overlay{/}{q}_2 - \overlay{/}{k}_1 + m_t \right) v(q_2) \nonumber\\
&&\times \left[ (p_1-p_2)_\rho \epsilon(p_1)\cdot \epsilon(p_2) +
2 p_2 \cdot \epsilon(p_1) \epsilon(p_2)_\rho -
2 p_1 \cdot \epsilon(p_2) \epsilon(p_1)_\rho \right ] \;,\\
i{\cal M}'_1 &=& i{\cal M}_1 \left[(p_1,a)\leftrightarrow (p_2,b)\right] \;,\\
i{\cal M}'_2 &=& i{\cal M}_2 \left[(p_1,a)\leftrightarrow (p_2,b)\right] \;,\\
i{\cal M}'_3 &=& i{\cal M}_3 \left[(p_1,a)\leftrightarrow (p_2,b)\right] \;,
\end{eqnarray}
where $ \epsilon(p_i)$'s denote the polarization 4-vectors for the gluons.

%%%%%%%%%%%%%%%%%%%%%%%%%%%%%%%%%%%%%%%%%%%%%%%%%%%%%

%%%%%%%%%%%%%%%%%%

\end{document}